# Comparative analysis of anisotropic material properties of uniaxial nematics formed by flexible dimers and rod-like monomers

Greta Cukrov[a†], Youssef Mosaddeghian Golestani[a†], Jie Xiang[a], Yu. A. Nastishin[a,b], Z. Ahmed[c], C. Welch[c], G. H. Mehl[c], and Oleg D. Lavrentovich[a*]

[a] Liquid Crystal Institute and Chemical Physics Interdisciplinary Program, Kent State University, Kent, OH 44242, USA

[b] Vlokh Institute of Physical Optics, 23 Dragomanov st., Lviv, 79005, Ukraine

[c] Department of Chemistry, University of Hull, Hull, HU6 7RX, UK

**Abstract**

We report temperature dependencies of material properties such as dielectric anisotropy, birefringence, splay ($K_{11}$), twist ($K_{22}$), and bend ($K_{33}$) elastic constants of the uniaxial nematic (N) phase formed by flexible dimers of DTC5C9 and compare their behavior to that of a corresponding monomer MCT5. DTC5C9 forms a twist-bend nematic ($N_{tb}$) at temperatures below the N phase. Anisotropic properties of MCT5 are typical of the rod-like mesogens. In particular, birefringence increases as the temperature is reduced, following the classic behavior, described by Haller. The elastic constants also follow the standard behavior, with their ratios being practically temperature-independent. In contrast, DTC5C9 shows a dramatic departure from the standard case. Birefringence changes non-monotonously with temperature, decreasing on approaching the N-$N_{tb}$ phase transition. $K_{33}$ decreases strongly to 0.4 pN near the N - $N_{tb}$ transition, although remains finite. The ratios of the elastic constants in DTC5C9 show a strong temperature dependence that can be associated with the bend-induced changes in the orientational distribution function. The measured elastic properties are consistent with the tendency of the dimeric molecules to adopt bent configurations that give rise to the $N_{tb}$ phase.

[*]Corresponding author: olavrent@kent.edu

[†]These authors contributed equally to the work

# 1. Introduction

In the past few years, dimeric liquid crystals (LCs), formed by molecules with two rigid rod-like units connected by a flexible bridge, attracted a strong research interest because of their fascinating features [1, 2, 3, 4, 5]. In particular, dimeric mesogens with an odd number of methylene flexible links form oblique helicoidal director structures in the nematic phases, in which the local director experiences bend and twist deformations. There are two experimentally observed realizations of the oblique helicoidal structures. First, the twist-bend nematic ($N_{tb}$) phase, exhibits the oblique helicoidal director with a nanometer pitch [3, 4] that exists as a ground state in a certain temperature range, below the standard uniaxial nematic (N) phase, in absence of any external fields and any chiral dopants [3, 4]. The second is a heliconical director state of a cholesteric liquid crystal, formed in the N phase doped with chiral molecules that is stabilized by an applied electric (or magnetic) field; as the external field changes, so does the heliconical pitch, usually on the micrometer length scales much larger than the molecular nanoscales [6, 7]. We denote this state as $Ch_{OH}$, where the subscript OH stands for "oblique helicoid". These two types of heliconical states represent yet another spectacular illustration of the subtle interplay between the details of molecular structure and macroscopic properties of LCs. This interplay has been envisioned in the theoretical predictions of the $N_{tb}$ [8, 9, 10] and field-induced oblique helicoidal structures [11]. In particular, Dozov [9] suggested that the twist-bend nematic could appear upon cooling of a uniaxial nematic formed by the molecules of a bent shape. The model [9] suggests that the bent shape can result in spontaneous bend deformation of the director and a change of the sign of the bend elastic constant $K_{33}$ from positive to negative as one crosses the N - $N_{tb}$ transition point ($T_{Ntb}$). Note that the transition might involve an appearance of other intermediate phases [12], but we do not discuss this possibility here as the focus is on the properties of the uniaxial

N phase. Uniform bend, however, cannot be realized in space without other types of deformations, either splay or twist. Thus two different variations of the nematic with spontaneous bend have been proposed by Meyer [8] and Dozov [9]: a twist-bend nematic and a splay-bend nematic (yet to be discovered experimentally). The relative stability of the two is controlled by the ratio of the splay $K_{11}$ to twist $K_{22}$ constants. Namely, in the twist-bend phase, $K_{11}/K_{22} > 2$, while in the splay-bend case, $K_{11}/K_{22} < 2$ [9, 10]. As for the Ch$_{OH}$ state, Meyer [11] predicted that it can be induced by an external electric or magnetic field in a cholesteric liquid crystal only when $K_{33} < K_{22}$. In calamitic nematics formed by rod-like molecules, the latter condition is not satisfied; the expected and universally observed trend follows the inequalities $K_{33} > K_{22} > K_{11}$ [13].

Prior research indicates that the dimeric materials feature elastic properties very different from those of rod-like nematogens. Atkinson et al. [14] demonstrated a strong odd-even effect in the behavior of $K_{33}$. Two dimers, one with an even number of alkyl groups in the spacer chain and another one with an odd number, were characterized by the splay Frederiks transition technique. In this approach, $K_{11}$ is determined from the threshold field of reorientation, and $K_{33}$ is obtained by extrapolating the voltage dependence of capacitance far above the threshold. $K_{33}$ of the odd homolog was about (3-4) pN, much smaller (by a factor $\sim 10$) than the corresponding value for the even homolog; the results were in good agreement with the theoretical consideration of the odd-even effect by Cestari et al [15].

The discovery of the N$_{tb}$ phase in the material abbreviated as CB7CB resulted in a strong interest in characterization of this dimer in its N phase. Yun et al [16] applied the extrapolation technique to extract $K_{33}$ from the splay Frederiks transition and found that in the temperature range (99.6-105) $^0$C, $K_{33}$ decreases as the material is cooled. $K_{33}$

is significantly smaller than both $K_{11}$ and $K_{22}$, reaching a minimum value of about 0.3 pN near the N- $N_{tb}$ transition. The ratio $K_{11}/K_{22}$ determined by Yun et al [16] was approximately ~ 1.7 near $T_{Ntb}$, i.e., lower than 2. The latter result is somewhat surprising, as $K_{11}/K_{22}$ is expected to be larger than 2 in CB7CB, since this material does exhibit the $N_{tb}$ phase rather than a splay-bend nematic, as demonstrated in the freeze-fracture transmission electron microscopy studies by Borshch et al [4] and by the resonant carbon soft X-ray scattering by Zhu et al [17]. Xiang et al. [6] explored the very same dimer CB7CB at the temperature 106 $^0$C and found that $K_{11}$ =5.7 pN and $K_{22}$ = 2.6 pN, i.e., $K_{11}/K_{22} \approx 2.2$, as expected; for CB7CB doped with 1wt % of the chiral dopant S811, $K_{33}$ was determined to be 0.3 pN. Finally, Parthasarathi et al [18] measured $K_{33}$ by the extrapolation approach similar to that of Yun et al [16] but arrived at a very different conclusion that $K_{33}$ of pure CB7CB is monotonically *increasing* as the sample is cooled down. It is only when CB7CB is mixed with a monomer 7OCB in some proportions that $K_{33}$ shows a non-monotonous behavior, first increasing and then decreasing as the temperature is lowered from the clearing point ($T_{NI}$).

Balachandran et al [19] used the splay Frederiks transition approach for another typical bimesogen, CB11CB, and found that as the temperature decreases, $K_{33}$ becomes smaller than $K_{11}$; namely, $K_{33}$ =6 pN, while $K_{11}$ =15.5 pN near $T_{Ntb}$. Adlem et al [20] explored multicomponent mixtures of dimeric materials by the dynamic light scattering; the latter allows one to extract all three bulk elastic constants of the material in the region close to the phase transition. They found even more dramatic difference, with $K_{33}$ ~ (0.3-2) pN and $K_{11}$ ~ (11-14) pN in the N phase at the temperature about one degree above $T_{Ntb}$; the ratio $K_{11}/K_{22}$ was determined to be somewhat higher than 2 [20].

All the experiments listed above were performed on materials with a positive dielectric anisotropy, $\Delta\varepsilon = \varepsilon_\parallel - \varepsilon_\perp > 0$, where the subscripts refer to the orientation of the director with respect to the field. The standard approach to measure $K_{11}$ and $K_{33}$ is to use a single uniformly aligned planar cell and apply the electric field across it. The threshold determines the splay constant $K_{11}$. As the field increases, the director acquires bend in addition to the predominant splay. By extrapolating and fitting the response function (such as capacitance of the cell or transmitted light intensity), one deduces $K_{33}$ [21]. The approach is very reasonable when $K_{33}$ is comparable to $K_{11}$. However, when $K_{33} \ll K_{11}$, one might wish to add an independent technique. A suitable direct method is a bend Frederiks transition caused by an external electric field in a so-called homeotropic cell, in which the director is perpendicular to the bounding plates and parallel to the applied field. To date, however, homeotropic alignment of dimeric nematics has been achieved only for the materials with negative dielectric anisotropy, $\Delta\varepsilon < 0$ [4]. Borshch et al [4] used homeotropic cells of a 7:3 mixture of DTC5C9 and MTC5, to measure $K_{33}$ directly. $K_{33}$ was much smaller than $K_{11}$, decreasing to 0.8 pN near $T_{Ntb}$.

The goal of this work is to determine the temperature dependencies of basic material properties, such as dielectric anisotropy $\Delta\varepsilon$, birefringence $\Delta n$, and elastic constants of splay $K_{11}$, twist $K_{22}$ and bend $K_{33}$ of the N phase formed by the dimeric molecules of DTC5C9 and to compare them to the corresponding values in the N phase formed by its monomer MCT5. All elastic constants are determined directly, by detecting the threshold of the Fredericks transitions for splay and twist in planar cells and for bend in homeotropic cells.

The results demonstrate that the monomer behavior is in line with the models based on rod-like building units, in which the ratios of the elastic constants such as $K_{11}/K_{33}$ and $K_{22}/K_{33}$ are practically temperature-independent, while all the constants grow as the temperature is lowered. In contrast, DTC5C9 shows a strong temperature dependence of $K_{11}/K_{33}$ and $K_{22}/K_{33}$. The ratio $K_{11}/K_{22}$ remains larger than 2 as expected for the material capable of forming the N$_{tb}$ phase as opposed to the splay-bend phase. Both $K_{11}$ and $K_{22}$ increase strongly when the temperature is lowered; $K_{33}$ exhibits a non-monotonous temperature dependence, first growing and then decreasing as the material is cooled down from $T_{NI}$. Near $T_{Ntb}$, $K_{33}$ reaches very small values, about 0.4 pN, that is more than one order of magnitude smaller than the corresponding value for MCT5.

## 2. Materials and methods

### *2.1 Chemical structure, phase diagram and alignment*

Figure 1 shows the chemical structures and the phase diagrams of the two studied materials, the monomer MCT5 (Figure 1a) and the dimer DTC5C9 (Figure 1b). The chemical formula of MCT5 is 2´,3´-difluoro-4,4´´-dipentyl-*p*-terphenyl (Kingston Chemicals Limited), while that one of DTC5C9 is 1,5-Bis(2´,3´-difluoro-4´´pentyl-[1,1´:4´1´´-terphenyl]-4-yl)nonane (synthesized in Hull). MCT5 shows N phase, while DTC5C9 exhibits N phase at high temperatures and N$_{tb}$ phase at lower temperatures. The presence of the N$_{tb}$ phase in DTC5C9 was established on the basis of the polarizing-microscopy textures showing the characteristic stripes and focal conic domains (Figure 2d), similar to those observed in other studies of the N$_{tb}$ phase [1, 4, 12, 22, 23]. DTC5C9 behavior mirrors broadly that of analogues with shorter and longer internal spacers and mesogens with negative dielectric anisotropy [24, 25, 26, 27, 28]. The clearing temperature of DTC5C9 increases in strong magnetic fields, with a rate ~0.6 ºC/T [29]; in all our experiments, we used magnetic field less than 1 T. Note that the formation of the $N_{tb}$ phase has been demonstrated by transmission electron microscopy for the 7:3 mixture of DTC5C9 and MCT5 [4].

To align both DTC5C9 and MCT5 homeotropically, we developed the following procedure. Glass substrates coated with transparent indium-tin-oxide (ITO) electrodes were treated with 1% water solution of Dimethyloctadecyl[3-(trimethoxysilyl)propyl]ammonium chloride (Sigma-Aldrich), on top of which a layer of polyimide aligning agent SE5661 (Nissan Chemical Industries) was spin-coated. The homeotropic alignment was stable in the entire N range, Figures 2a,b and 3a,b. Planar alignment was achieved by using rubbed polyimide PI2555 layers (HD Microsystems), Figure 2c,d and 3c,d. On cooling the homeotropic and planar cells, the N - $N_{tb}$ transition is clearly evidenced by a propagating front that quenches fluctuations of the director in the N phase. All textural observations were performed using a polarizing optical microscope (POM) [Nikon, Optiphot-2 Pol] equipped with a home-made hot stage with the temperature control accuracy 0.1 °C. The cells thickness $d$ was set by silicon microspheres of a calibrated diameter dispersed in a UV-curable glue; $d$ was measured with a light interference technique using a spectrometer.

*2.2 Dielectric anisotropy*

For dielectric characterization, we used LCR meter (HP4284A) to measure the capacitance of the cells. The dielectric measurements of MCT5 were performed using a planar cell with $d = 19.3 \pm 0.1\,\mu m$; and a homeotropic cell with $d = 19.5 \pm 0.2\,\mu m$. For the measurements of DTC5C9, we used a planar cells with $d = 21.5 \pm 0.2\,\mu m$; and a homeotropic cell with $d = 37.6 \pm 0.2$ μm. Dielectric permittivities were measured using an AC electric field at frequencies $f = 40$ kHz for MCT5 and $f = 10$ kHz for DTC5C9. The relatively high frequency of the field assured that the flexoelectric and surface polarization contributions [30] to the electro-optical response of the cell are minimized. The dielectric permittivity $\varepsilon_\parallel$ was determined from the capacitance of the homeotropic cells, while $\varepsilon_\perp$ was measured using the planar cells; $\Delta\varepsilon = \varepsilon_\parallel - \varepsilon_\perp$ is negative in the entire N range for both materials, Figure 5.

*2.3 Birefringence*

The temperature dependence of the birefringence $\Delta n$ for MCT5 and DTC5C9 was determined by measuring the optical retardance, $\Gamma = \Delta n d$, on cooling at a rate of 2.0ºC/min using LC PolScope (Abrio Imaging System) [31, 32], Figure 6. The sample was probed with a monochromatic light of wavelength $\lambda = 546$ nm. The accuracy of the retardance measurements is not worse than 1 nm. The measurements of $\Delta n$ were performed using planar cells of thicknesses $d = 3.56 \pm 0.05$ μm for MCT5 and $d = 5.09 \pm 0.05$ μm for DTC5C9.

*2.4 Elastic constants*

The elastic constant of splay $K_{11}$ was obtained by determining the magnetic threshold, $B_{th1}$, of the splay Frederiks transition in a planar sample caused by the magnetic field applied perpendicularly to the bounding plates. The director reorientation was probed by measuring the transmitted light intensity passing normally through the LC cell between crossed polarizers. $K_{11}$ was determined from the relationship

$$K_{11} = \left(\frac{dB_{th1}}{\pi}\right)^2 \frac{\Delta\chi}{\mu_o}, \tag{1}$$

where $B_{th1}$ is the magnetic threshold, $\Delta\chi$ is the diamagnetic anisotropy, and $\mu_o = 4\pi \times 10^{-7}$ H·m$^{-1}$.

There are several approaches to measure $K_{22}$, such as the Frederiks transition in a twisted nematic (TN) cell [33], in-plane realignment in a planar cell [34], a dynamic light scattering method [35], etc. The measured value of $K_{22}$ in these methods is usually less accurate than those of $K_{11}$ and $K_{33}$, since the twist deformations are accompanied by bend and splay. For example, in the TN method, $K_{22}$ depends on $K_{11}$ and $K_{33}$, so that the errors in $K_{11}$ and $K_{33}$ are accumulated into the value of $K_{22}$. In the in-plane switching cell method, $K_{22}$ is independent of $K_{11}$ and $K_{33}$, but the electric field within the cell is not uniform, resulting in an additional source of errors [36]. To avoid these complications, we use the twist Frederiks transition caused by the magnetic field acting on a planar cell in such a way that the field is perpendicular to the director but lies in the plane of the cell. At a certain threshold, $B_{th2}$, the uniform cell experiences pure twist deformation, which allows us to determine the twist elastic constant as

$$K_{22} = \left(\frac{dB_{th2}}{\pi}\right)^2 \frac{\Delta\chi}{\mu_o}. \tag{2}$$

We used a planar sample placed between two crossed polarizers and tested transmittance of a He-Ne laser beam ($\lambda = 632.8$ nm). The beam was directed at an angle large enough to overcome the Mauguin regime [37, 38]. The output light intensity was measured as a function of the magnetic field to determine $B_{th2}$ and thus $K_{22}$.

In order to obtain $K_{33}$, we applied an AC electric field using LCR meter and measured the voltage dependence of the transmitted light passing through a homeotropic cell. We applied 40 kHz for MCT5; 10 and 40 kHz for DTC5C9 measurements. The bend Frederiks threshold, $V_{th3}$, was determined from the optical phase retardation vs voltage curve using the so-called double extrapolation approach [39, 40], from which the value of $K_{33}$ was extracted,

$$K_{33} = \frac{\varepsilon_o \Delta \varepsilon V_{th3}^2}{\pi^2}. \tag{3}$$

Figure 4 illustrates the typical field dependence of optical phase retardation used to find $V_{th3}$ by double extrapolation. We also employed the capacitance method [41] for determining the bend elastic constant using the capacitance vs voltage dependency measured for a homeotropic cell.

The bend Frederiks transition in a homeotropic cell can also be triggered by the magnetic field directed perpendicularly to the director, as $\Delta\chi$ of MCT5 and DTC5C9 is positive, $\Delta\chi = \chi_\parallel - \chi_\perp > 0$. The magnetic threshold $B_{th3}$ was determined by measuring the transmitted light intensity passing through the homeotropic LC cell. Using the expression

$$K_{33} = \left(\frac{dB_{th3}}{\pi}\right)^2 \frac{\Delta\chi}{\mu_o} \tag{4}$$

and comparing the bend constant to the value obtained in the electric Frederiks effect, we determined the diamagnetic anisotropy, $\Delta\chi = \varepsilon_o \mu_o \Delta\varepsilon \left(\frac{V_{th3}}{dB_{th3}}\right)^2$. For example, $\Delta \chi = (1.1 \pm 0.1) \times 10^{-6}$ (SI units) at $T - T_{NI}$ = -22 °C for MCT5, and $\Delta \chi = (1.4 \pm 0.1) \times 10^{-6}$ (SI units) at $T - T_{NI}$ = -25 °C for DTC5C9. The measured values of $\Delta \chi$ are used in measuring the splay and twist elastic constants in planar cells.

We used two cells of different thicknesses (19.4 μm and 37.6 μm) with the same homeotropic alignment layer and found the same value of $B_{th}d$ for both cells. The latter justifies that the polar anchoring of the homeotropic alignment is strong enough and that the threshold field and the $K_{33}$ data are not affected by the finite surface anchoring. The threshold field values in other geometries were determined in a similar fashion.

## 3. Results

### *3.1 Dielectric anisotropy*

Both MCT5 and DTC5C9 liquid crystals show a negative dielectric anisotropy, Figure 5. MCT5 exhibits a monotonous temperature behavior that agrees with data collected by Urban et al for the same compound (referred to as KS3) [42]. Behavior of DTC5C9 is dramatically different near the N-N$_{tb}$ transition: the absolute value of the dielectric anisotropy starts to decrease in the pretransitional region, similarly to the data by Borshch et al obtained for the 7:3 mixture of DTC5C9 and MCT5 [4].

### *3.2 Birefringence*

Birefringence of MCT5 increases monotonically as the temperature is reduced, in a fashion typical for rod like molecules [43], following the Haller's rule [44] in the entire nematic range, Figure 6a. Haller's rule is of the form

$$\Delta n = \delta n \left(1 - \frac{T}{T^*}\right)^{\beta}, \tag{5}$$

where $\delta n$, $T^*$, and $\beta$ are the fitting parameters. For MCT5 we find $\delta n = 0.309 \pm 0.003$, $T^* = 390.3 \pm 0.2$ K and $\beta = 0.178 \pm 0.004$. As suggested by Geppi et al [45], validity of the fit in Equation 5 allows one to approximate the temperature dependence of the order parameter as $S(T) = \frac{\Delta n(T)}{\delta n}$. Using this approximation, we find the value of $S$ increasing from 0.4 to 0.7 within the entire N phase, Figure 7, in agreement with the data presented by Geppi et al [45] for the compounds very similar to MCT5, such as KS7, in which one of the aliphatic tails is shorter than that of the aliphatic tails of MCT5 by one methyl group.

In contrast to the monotonous behavior of $\Delta n$ in MCT5, the birefringence of DTC5C9 cooled down from the isotropic phase first increases and then decreases on approaching $T_{Ntb}$, Figure 6b. The decrease of $\Delta n$ near the $N_{tb}$ phase is consistent with other studies of dimers [4, 46, 47]. The behavior of $\Delta n$ does not obey Haller's rule, clearly deviating near the N-$N_{tb}$ transition. The temperature range of this deviation is broad, about 10 K. For this reason we use the Haller fit only for the data above $T = T_{Ntb} + 10$ K. The result is shown in Figure 6b, yielding $\delta n = 0.219 \pm 0.006$, $T^* = 434.6 \pm 0.4$ K and $\beta = 0.140 \pm 0.009$. Non-monotonous $\Delta n(T)$ behavior suggests that the order parameter $S(T)$ might also be non-monotonous. As a rough approximation, we assume that the two functions are connected in a linear fashion, $S(T) = \frac{\Delta n(T)}{\delta n}$, where $\delta n = 0.219$ is the value obtained from the fit above. More accurate measurement of $S(T)$ would require independent measurements of ordinary and extraordinary refractive indices. The dependence $S(T)$ calculated from the entire range of data on $\Delta n(T)$, is non-monotonous, with a maximum $S_{max} = 0.68$ achieved at about 10 K above $T_{Ntb}$, Figure 6b.

Interestingly, the non-monotonous $S(T)$ with the maximum at $T_{Ntb}+10$ K has been already reported by Emsley et al [48] who measured the order parameter from chemical shift anisotropies in NMR experiments. A similar conclusion about pretransitional changes of $S(T)$ that start at $T_{Ntb}+10$ K were made by Burnell et al [49] for the mixture of flexible dimer CB9CB with 5CB.

*3.3 Elastic constants*

The elastic constants of MCT5 (Figure 8) follow general behavior of calamitic LCs, as $K_{33} > K_{11} > K_{22}$ [19, 50]. The temperature behavior of the elastic constants of DTC5C9 is dramatically different in comparison to MCT5. The ratio $K_{11}/K_{22} > 2$ (Figure 9a), as expected by Dozov for materials exhibiting the N$_{tb}$ phase [9]. On departure from $T_{NI}$, $K_{33}$ first increases and then dramatically decreases on approaching $T_{Ntb}$ (Figure 10).

**4. Discussion**

In conventional rod-like nematic liquid crystals (such as MCT5), far from the phase transitions into lower-symmetry phases, the anisotropic properties such as birefringence and dielectric anisotropy change linearly with the scalar order parameter. The birefringence obeys the Haller type behavior, Figure 6, implying that the scalar order parameter of MCT5 monotonously increases as the temperature is lowered from $T_{NI}$. In the simplest mean-field models, the elastic constants are expected to follow the relationship $K_{ii} \propto S^2$, while showing the inequalities $K_{33} > K_{11} > K_{22}$ [13, 50]. The temperature behavior of the elastic constants for MCT5 follows these expectations rather well. Indeed, Figure 8 demonstrates that $K_{33} > K_{11} > K_{22}$. The ratio of any two elastic constants is roughly independent of temperature, as shown for MCT5 by open symbols in Figure 9.

DTC5C9 shows a dramatically different behavior. First of all, $\Delta n$ shows a non-monotonous temperature dependence, decreasing near N-N$_{tb}$ transition, within a rather broad temperature range. The maximum $\Delta n$ is achieved at the temperature ~10 K above $T_{Ntb}$, suggesting that $S$ follows a similar non-monotonous behavior with a maximum at a similar temperature. NMR measurements by Emsley et al [48] for DCT5C9 show that $S(T)$ is indeed reaching a maximum at 10 K above $T_{Ntb}$. The non-monotonous temperature dependencies of $\Delta n$ and $S$ could be associated with the growth of the population of molecular conformers with a bent shape at lower temperatures and higher densities. It is very likely that these bent molecules form pretransitional clusters with the structure closely resembling the structure of the N$_{tb}$ phase. We remind here that according to the theory, the driving force of the N$_{tb}$ phase formation is the tendency of molecules to bent; this bend is uniform in only when accompanied by twist [8, 9]. The twist-bend distortions set up a one-dimensional periodic modulation of the local director. In the pretransitional region, one would expect a similar interplay of spontaneous bend and twist so that the pretransitional clusters could develop some pseudo-layered structure.

We turn now to the discussion of the elastic properties of DTC5C9. The splay elastic constant is noticeably larger than that in MCT5, Figure 8. For example, at the temperature $T - T_{NI} = -10$ °C, MTC5 and DTC5C9 show $K_{11} = 4$ and 7 pN, respectively, while at $T - T_{NI} = -20$ °C, the values are $K_{11} = 7$ and 10 pN. This result correlates with the model proposed by Meyer, in which $K_{11}$ is expected to increase linearly with the length $L$ of the molecules [51], $K_{11} = \frac{k_B T}{4d} \frac{L}{d}$, where $k_B$ is the Boltzmann constant, $T$ is the absolute temperature, $d$ is the diameter of the rod-like molecule. Splay deformations require creation of gaps between the molecules. To keep the density constant, these gaps must be filled with the ends of adjacent molecules. The formula above follows from the consideration of the entropy loss associated with rearrangements of the molecular ends assumed to behave as non-interacting particles of an ideal gas [51]. Such an assumption is well justified for long molecules, but might be less accurate for relatively short ones. It is worth noticing that other monomer-dimer comparative studies do not show such a large differences in the values of $K_{11}$ [52] and sometimes even show that $K_{11}$ of dimers is smaller than $K_{11}$ of a monomer. Dilisi et al [53] measured $K_{11}$ for a monomer 4,4'-dialkoxyphenylbenzoate [$C_5H_{11}OC_6H_4COOC_6H_4OC_5H_{11}$], its related "even" dimer with a spacer of ten methylene groups [52] and an "odd" dimer with nine methylene groups in the flexible bridge [53]. It turned out that the odd dimer, presumably of a bent shape similar to that one of DTC5C9, produced the lowest $K_{11}$ among the three studied molecules [52, 53], contrary to our case. For example, at $T - T_{NI} = -10\ ^{\circ}C$, the values of $K_{11}$ for the monomer, even dimer, and odd dimer, are 7.5, 8.5, and 5 pN, respectively [52, 53]. One might argue that a reduced scalar order parameter of the odd dimer (which is indeed evidenced by the lower diamagnetic anisotropy [53]) might lead to a smaller $K_{11}$. However, this mechanism is apparently not the main one governing $K_{11}$ in DTC5C9 and MCT5, and the increase of the molecular length appears to be dominating.

The twist elastic constants of DTC5C9 and MCT5 show similar values in the upper temperature range but behave differently at lower temperatures. Namely, $K_{22}$ of DTC5C9 increases noticeably near the transition into the N$_{tb}$ phase. An increase of $K_{22}$ is typical near the nematic – smectic A phase transition phase and is explained by the formation of cybotactic clusters with periodic structure of equidistant layers that hinder deformations of twist and bend of the normal to the layers. A similar mechanism should be expected in the vicinity of $T_{Ntb}$, even if the smectic order, in the sense of periodic mass density modulation, does not develop. The $N_{tb}$ structure is one-dimensionally periodic because of twist-bend director modulation and thus also prohibits deformations of twist and bend of the heliconical axis. The increase of $K_{22}$ can thus be treated as a pretransitional effect potentially caused by the formation of pretransitional clusters with a pseudo-layered structure. Such an effect correlates well with the pretransitional decrease of birefringence and scalar order parameter in DCT5C9 discussed above. The measured values of $K_{11}$ and $K_{22}$ show that $K_{11}/K_{22} > 2$; the result agrees with the Dozov model of the $N_{tb}$ phase [9].

The most spectacular deviation of DTC5C9 from the classic picture of nematic elasticity is demonstrated by $K_{33}$, Figure 10. As the temperature decreases from $T_{NI}$, $K_{33}$ first increases, but after reaching a maximum, $K_{33}$ decreases to a very low value of 0.4 pN. The rate of the decrease is slowed down near the N-$N_{tb}$ transition; there seems to be a small plateau or even an increase right before the transition. Such a behavior has been already observed in other studies and is associated with the bent conformations of the dimeric molecules [20, 46]. Importantly, the temperature dependence of $K_{33}$ in DTC5C9 does not correlate with the expected behavior of the scalar order parameter, as clearly evidenced by a strong temperature dependencies of the ratios $K_{22}/K_{33}$ and $K_{11}/K_{33}$, Figure 9b,c.

The idea that bent shape of molecules can lead to a small bend constant $K_{33}$ dates back at least four decades to the theoretical works by Gruler [54] and Helfrich [55]. The underlying mechanism is that in presence of bend deformations, the distribution of molecular orientations is no longer axially symmetric; the molecules can realign cooperatively, mimicking the director bend by adjusting their bend shapes and thus relieving the elastic strain. A reduction of $K_{33}$ through bent conformations has also been predicted by Terentjev and Petshchek [56] for semiflexible molecules. Experimentally, a reduction in $K_{33}$ was indeed confirmed for mixtures [57] and pure compounds with rigid bent core molecules [58, 59, 60].

The considerations by Gruler and Helfrich have been recently reinforced in the density functional theory by De Gregorio et al [61]. Namely, mesogens of bent shape are shown to exhibit a non-monotonous dependence of $K_{33}$ on the number density of the molecules. As a function of an increasing $S$, $K_{33}$ is predicted [61] to first increase (near $T_{NI}$), then reach a maximum and decrease to 0, presumably when the material undergoes a transition into the N$_{tb}$ phase. Importantly, the effect of a reduced $K_{33}$ is observed only if the bent-core molecules are collectively adjusting to the imposed bend. If the orientational distribution of the bent region remains locally uniaxial, the model [61] predicts a strong increase of $K_{33}$ with the enhancement of $S$, a classic result for the rod-like mesogens.

It is important to stress that the experimental values of $K_{33}$, unlike some of their theoretical counterparts, never reach zero at $T_{Ntb}$ and might even slightly increase near the transition [20]. The most likely reason is formation of pretransitional clusters with local bend and twists which establish a pseudo-layered structure, similarly to the discussed case of an increased $K_{22}$.

Finally, $K_{33} < K_{22}$ found for DTC5C9 opens a possibility for realization of the Ch$_{OH}$ cholesteric under an applied electric field as predicted by Meyer [11] making the cholesteric pitch and thereby the selective reflection band tuneable [6, 7].

## 5. Conclusion

We determined the material parameters of the uniaxial nematic phase formed by a monomer MCT5 and the flexible odd dimer DTC5C9. The monomer shows a classic temperature dependence of the material properties expected for rigid rod-like mesogens. The dimer DTC5C9 exhibits a dramatic departure from this behavior. First, the birefringence of DTC5C9 does not follow the Haller's rule in a broad temperature range of width ~10 K near $T_{Ntb}$. This behavior correlates well with the idea that the scalar order parameter in the pretransitional region decreases, because of the growth of population of molecules with bent conformations and formation of pretransitional clusters with local bends and twists. The twist elastic constant increases near the N-$N_{tb}$ transition, while the bend constant decreases to about 0.4 pN. The ratios of the elastic constants such as $K_{11}/K_{33}$ and $K_{22}/K_{33}$ show a strong temperature dependence, emphasizing that $K_{33}$ does not correlate with the scalar order parameter. It is very likely that the bend deformations used to directly determine $K_{33}$ in the bend Frederiks transition are relieved by formation of polar biaxial structures with the bent molecules packed in a similar fashion, thus modifying the orientational distribution function. Finite values of $K_{33}$ and an increase of $K_{22}$ near the transition into the $N_{tb}$ phase can be associated with the pretransitional formation of pseudo-layered clusters.

**Acknowledgements**

We thank S. V. Shiyanovskii, S. Zhou, Y. K. Kim, S. Paladugu and C. Peng for useful discussions.

**Funding**

This work was supported by NSF grant [DMR-1410378]. ZA and CW acknowledge funding through the EPSRC projects [EP/M015726] and [EP/J004480].

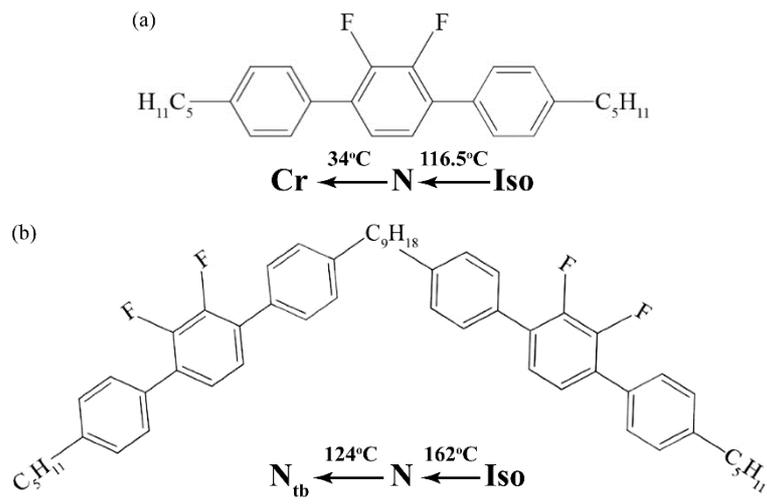

**Figure 1.** Chemical structures and phase diagrams of the liquid crystal monomer MCT5 (a) and dimer DTC5C9 (b) with negative dielectric anisotropy. The phase diagram was determined on cooling at the rate of 0.1°C/min from the isotropic phase.

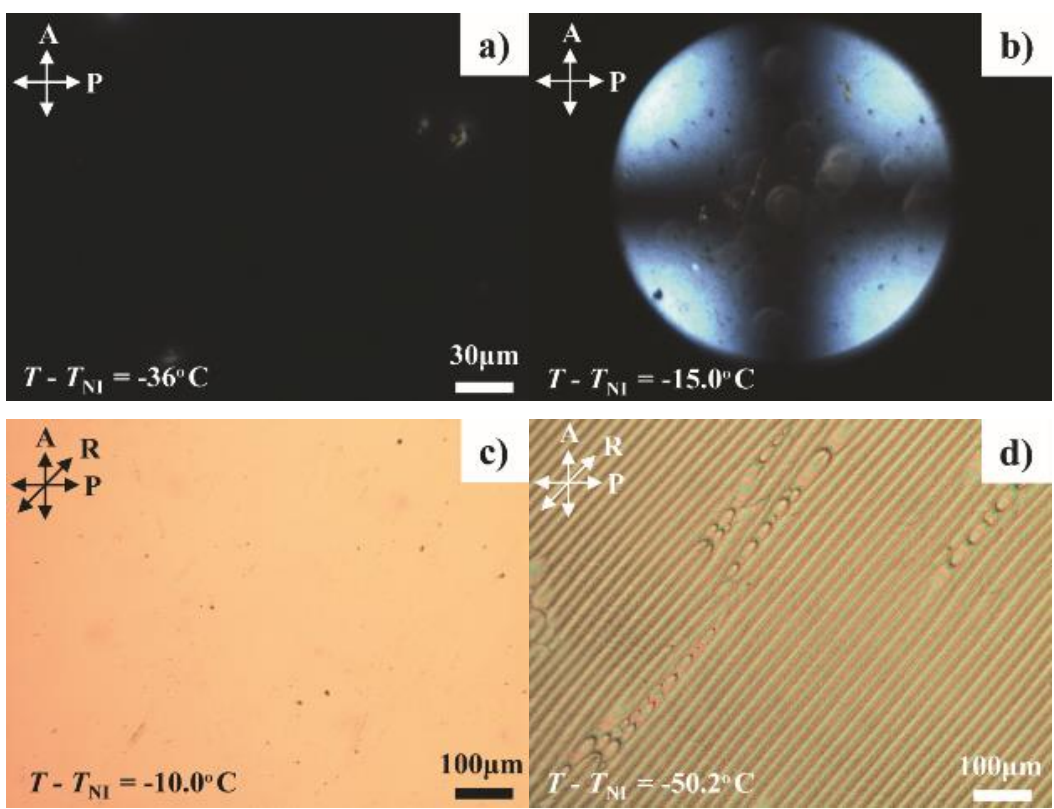

**Figure 2.** Polarizing optical microscope textures of DTC5C9 in the (a,b) homeotropic N cell ($d = 19.4\,\mu m$) and (c) planar N cell ($d = 21.5\,\mu m$); (d) planar $N_{tb}$ cell with stripes and focal conic domains ($d = 21.5\,\mu m$). Part (b) shows the conoscopic pattern characteristic of a homeotropic uniaxial nematic. The director in part (c) is along the rubbing direction R; polarizer and analyzer are labelled as P and A.

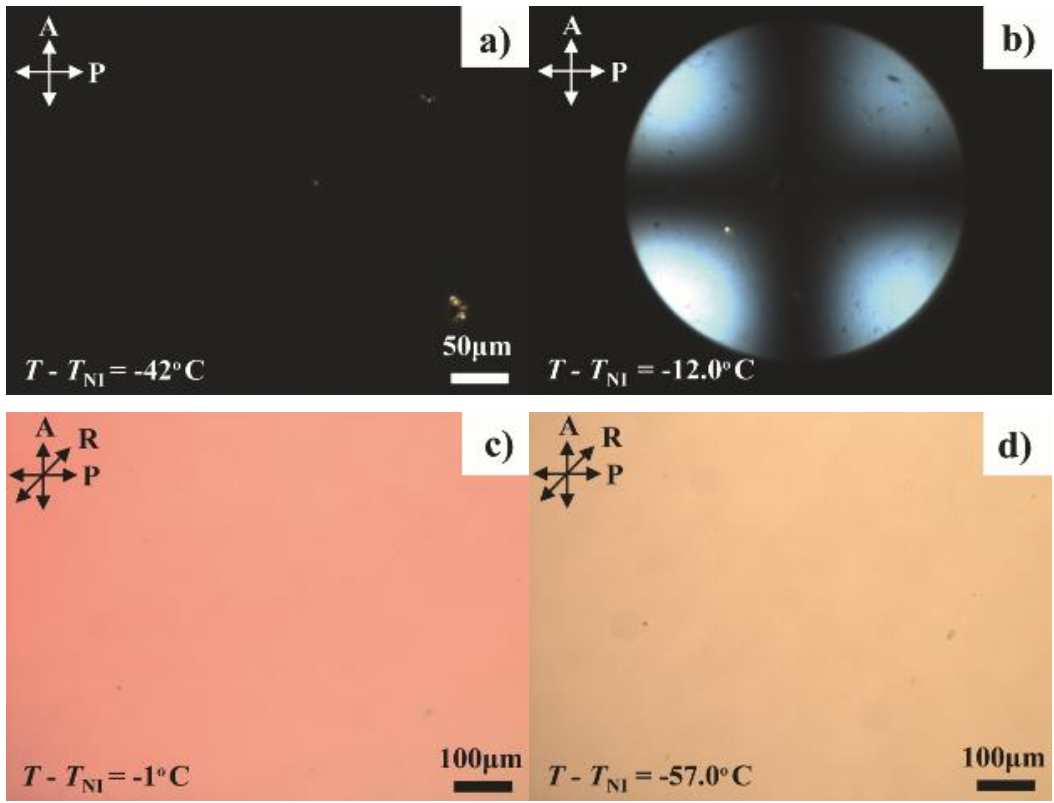

**Figure 3.** Polarizing optical microscope textures of MCT5 in the (a,b) homeotropic N cell (cell thickness $d = 19.5\,\mu m$); (c,d) planar N cell ($d = 19.5\,\mu m$). Part (b) shows the conoscopic pattern characteristic of a homeotropic uniaxial nematic. The director in part (c) and (d) is along the rubbing direction R.

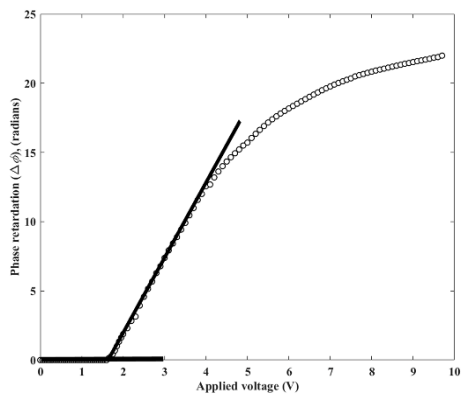

**Figure 4.** Determination of $V_{th3}$ for DTC5C9 material in a homeotropic cell with $d = 19.4\,\mu m$ at $T - T_{NI} = -24^\circ C$. The bold straight lines illustrate how the threshold is determined by double extrapolation.

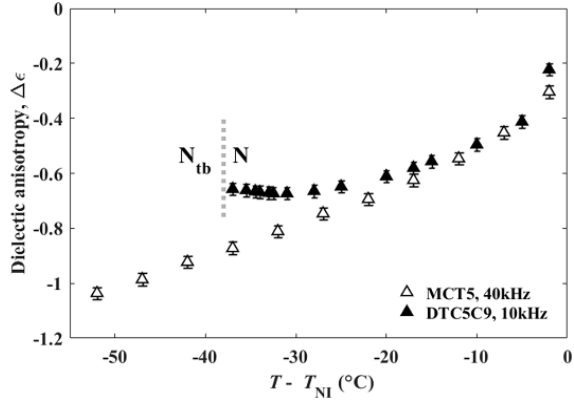

**Figure 5.** Temperature dependence of the dielectric anisotropy for liquid crystal monomer MCT5 (open symbols) and dimer DTC5C9 (filled symbols) measured at frequencies of 40 and 10 kHz respectively. The dashed vertical line represents N-$N_{tb}$ transition temperature for DTC5C9.

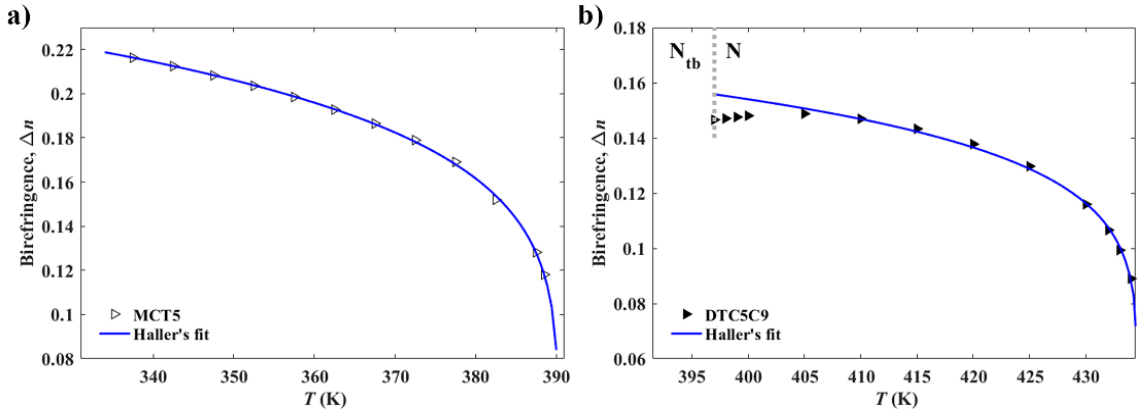

**Figure 6.** Temperature dependence of $\Delta n$ for (a) monomer MCT5 (open symbols) and (b) dimer DTC5C9 (filled symbols). The wavelength of the probing light was 546 nm.

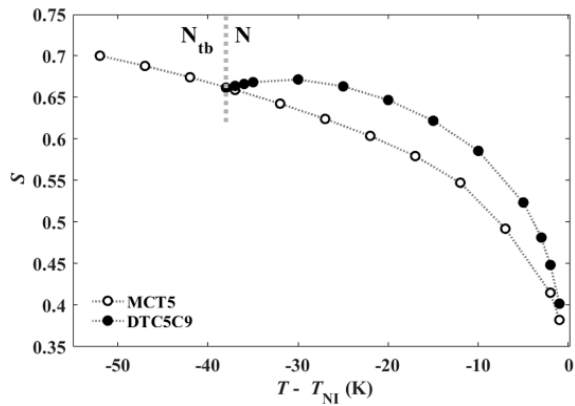

**Figure 7.** Temperature dependence of $S$ for the monomer MCT5 (open symbols) and dimer DTC5C9 (filled symbols). Dotted lines connecting the data points are guides for eyes.

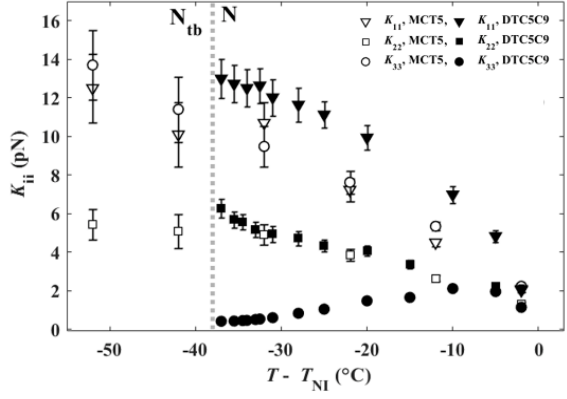

**Figure 8.** Temperature dependent elastic constants of MCT5 (open symbols) and dimer DTC5C9 (filled symbols).

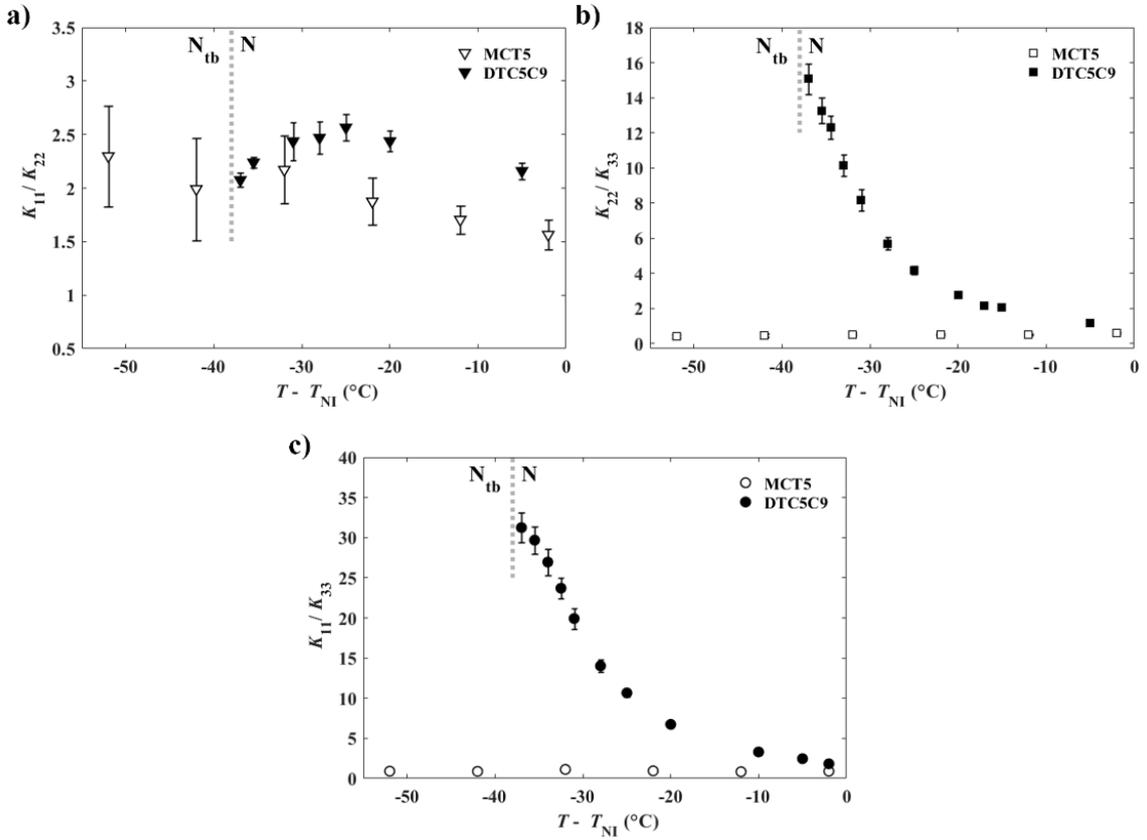

**Figure 9.** Temperature dependencies of the ratios (a) $K_{11}/K_{22}$ (b) $K_{22}/K_{33}$ and (c) $K_{11}/K_{33}$ for MCT5 (open symbols) and DTC5C9 (filled symbols).

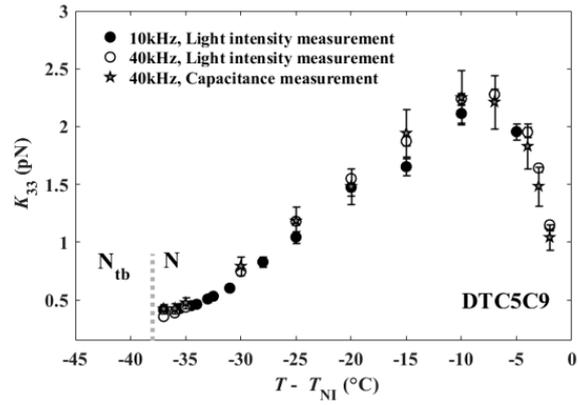

**Figure 10.** Temperature dependence of $K_{33}$ for DTC5C9. Filled and empty circles represent the results obtained by the light intensity measurements with applied electric fields at 10 and 40 kHz respectively. The stars represent $K_{33}$ extrapolated from the capacitance vs voltage curve.